\documentclass{amsart}

\usepackage{amssymb,amscd,amsthm,amsxtra}
\usepackage{latexsym}
\usepackage{amsmath,amsfonts}

\def\Datum{February 29, 2012}

\newtheorem{theorem}{Theorem}[section]

\theoremstyle{definition}

\theoremstyle{remark}

\numberwithin{equation}{section}

\def\tu{\widetilde{u}}
\def\ux{{\underline{x}}}

\def\N{{\mathbb N}}

\def\alphaset{{\mathfrak A}}

\def\Gspace{{\mathfrak G}}

\def\tr{{\rm Tr}}

\def\bra{\big\langle}
\def\ket{\big\rangle}

\def\cH{{\mathcal H}}

\def\tu{\widetilde{u}}
\def\ux{{\underline{x}}}

\def\eqnn{\begin{eqnarray*}}
\def\eeqnn{\end{eqnarray*}}
\def\eqn{\begin{eqnarray}}
\def\eeqn{\end{eqnarray}}

\newcommand{\comment}[1]{}
 
\newcommand{\pj}{\nabla_{j}}
\newcommand{\pk}{\nabla_{k}}
\newcommand{\pkup}{\nabla^{k}}
\newcommand{\pjup}{\nabla^{j}}
\newcommand{\R}{\mathbb R}
\newcommand{\C}{\mathbb C}

\begin{document}

\title[Morawetz identities for GP hierarchy]{Multilinear Morawetz identities
for the Gross-Pitaevskii hierarchy}
\author{T. Chen}
\address{Department of Mathematics, University of Texas, Austin, TX, 78712, USA.}
\email{\tt tc@math.utexas.edu}

\author{N. Pavlovi\'{c}}
\address{Department of Mathematics, University of Texas, Austin, TX, 78712, USA.}
\email{\tt natasa@math.utexas.edu}

\author{N. Tzirakis}
\address{Department of Mathematics, University of Illinois at Urbana-Champaign, Urbana, IL, 61801, USA.} \email{\tt tzirakis@math.uiuc.edu}

\date{\Datum}


\begin{abstract}
This article consists of two parts. In the first part, we review the most recent proofs establishing quadratic Morawetz inequalities for the nonlinear Schr\"odinger equation (NLS). We also describe the applications of these estimates to the problem of quantum scattering. In the second part, we
generalize some of the methods developed for the NLS by many authors
to the case of  Gross-Pitaevskii (GP) hierarchies.
In particular, we prove both regular and interaction
Morawetz identities for the GP hierarchy, which appear here for the first time in the literature.
\end{abstract}
\maketitle

\section{Introduction}
In this paper, we discuss some recent a priori estimates for the solutions to the power type nonlinear Schr\"odinger equation
\begin{equation}\label{nls}
\left\{
\begin{matrix}
iu_{t}+\Delta u =\lambda|u|^{p-1}u, & x \in {\mathbb R^n}, & t\in {\mathbb R},\\
u(x,0)=u_{0}(x)\in H^{s}({\mathbb R^n})
\end{matrix}
\right.
\end{equation}
for any $p>1$ and $\lambda \in \R$. This involves the proof of Morawetz identities and
Morawetz inequalities for the NLS.
Moreover,   we derive  generalizations of both regular (one-particle), and
interaction Morawetz identities for Gross-Pitaevskii hierarchies.

Gross-Pitaevskii  (GP) hierarchies are infinite systems of coupled linear non-homogeneous
PDEs that appear naturally in the derivation of the nonlinear Schr\"odinger equation,
as the dynamical mean field limit of the manybody quantum dynamics of  Bose gases
(under the Gross-Pitaevskii scaling).
Let $q\in\{2,4\}$ and $\lambda \in \{-1, 1\}$. The $q$-GP hierarchy on $\R^d$
is an infinite system of coupled PDEs for functions (referred to as {\em marginal density matrices})
$\gamma^{(k)}(t, x_1, \dots, x_k, x'_1, \dots, x'_k)$
for $k\in\N$, $t \in \R$ and $x_i, x'_i \in \R^d$ for $i \in \{1, \dots, k\}$:
\eqn \label{intro-GP}
    i\partial_t \gamma^{(k)} \, = \, -\sum_{j=1}^k \left( \Delta_{x_j} - \Delta_{x'_j} \right) \gamma^{(k)}
    \, + \,  \lambda B_{k+\frac q2} \gamma^{(k+\frac q2)},
\eeqn
where the inhomogeneous term is given by
\eqn \label{eq-def-b}
    B_{k+\frac q2}\gamma^{(k+\frac{q}{2})}
    \, = \,  \sum_{j=1}^k \left( B^+_{j;k+1,\dots,k+\frac q2}\gamma^{(k+\frac{q}{2})}
    - B^-_{j;k+1,\dots,k+\frac q2}\gamma^{(k+\frac{q}{2})} \right)\,.
\eeqn
with
\begin{align*}
& \left( B^+_{j;k+1,\dots,k+\frac q2}\gamma^{(k+\frac{q}{2})} \right)(t, x_1, \dots, x_k, x'_1, \dots, x'_k) \\
& \quad \quad  = \gamma^{(k+\frac q2)}(t,x_1,\dots,{\bf{x_j}}, \dots, x_k,  {\underbrace{ \bf{ x_j, \cdots x_j }}_{\frac q2} };x_1',\dots,x_k', {\underbrace{\bf{ x_j, \cdots, x_j }}_{\frac q2} }),
\end{align*}

\begin{align*}
& \left( B^-_{j;k+1,\dots,k+\frac q2}\gamma^{(k+\frac{q}{2})} \right)(t, x_1, \dots, x_k, x'_1, \dots, x'_k) \\
& \quad \quad =  \gamma^{(k+\frac q2)}(t,x_1,\dots, x_k,  {\underbrace{\bf{ x_j', \cdots x_j' }}_{\frac q2} };x_1',\dots,{\bf{x_j'}}, \dots x_k', {\underbrace{\bf{ x_j', \cdots, x_j' }}_{\frac q2} }) \,,
\end{align*}
accounting for the interactions between $1+\frac q2$ particles.
The marginal density matrices are hermitean,
$\gamma^{(k)}(\ux_k, \ux'_k) = \overline{\gamma^{(k)}(\ux'_k,\ux_k)}$, and
satisfy the property of {\em admissibility}, that is,
\eqn\label{eq-admiss-def-1}
	\gamma^{(k)}(t,x_1,\dots,x_k;x_1',\dots,x_k')
	\, = \,
	\int dy \, \gamma^{(k+1)}(t,x_1,\dots,x_k,y;x_1',\dots,x_k',y)
\eeqn
for all $k\in\N$, with
normalization $\int dx \,\gamma^{(1)}(t,x;x)=1$.

\subsection{Nonlinear Schr\"{o}dinger equation}

Equation \eqref{nls} is the Euler-Lagrange equation for the Lagrangian density
$$L(u)=-\frac{1}{2}\Delta(|u|^2)+\lambda\frac{p-1}{p+1}|u|^{p+1}.$$
Several invariants of the equation lead to conservation laws that play a central role in the well-posedness theory of the problem. The homogeneous Sobolev norm $\dot{H}^{s_c}$ is invariant under the scaling $u^{\mu}(x,t)=\mu^{-\frac{2}{p-1}}u(\frac{x}{\mu},\frac{t}{\mu^2})$, when $s_c=\frac{n}{2}-\frac{2}{p-1}$. This critical regularity already partitions the general problem into different sub-problems with a varied degree of difficulty. Two important cases that are well understood correspond to the energy subcritical problem for $1<p<1+\frac{4}{n-2},\ n\geq 3,\ 1<p<\infty,\ n=1,2$ and initial data in $H^1$, and the $L^2$ subcritical problem for $1<p<1+\frac{4}{n},\ n\geq 1$ and $L^2$ initial data. Time translation
invariance leads to energy conservation
\begin{equation}\label{energy}
E(u)(t)=\frac{1}{2}\int |\nabla u(t)|^{2}dx+\frac{\lambda}{p+1}\int |u(t)|^{p+1}dx=E(u_{0}).
\end{equation}
In the defocusing case, ($\lambda=1$), this law provides an $H^1$ a priori bound that extends any local energy subcritical solution to an interval of arbitrary length. The equation is also invariant under phase rotation which leads to the conservation of mass law
\begin{equation}\label{mass}
\|u(t)\|_{L^{2}}=\|u_{0}\|_{L^{2}}.
\end{equation}
This implies for any $\lambda$ the existence of global-in-time solutions for solutions evolving from finite mass data in the $L^2$-subcritical case of $p<1+\frac{4}{n}$. For the $L^2-$critical problem, $p=1+\frac{4}{n}$, the time of the local theory depends on the profile of the initial data, in addition to their $L^2$ norm, and thus the conservation law is not immediately useful.
Space translation invariance leads to momentum conservation
\begin{equation}\label{momentum}
\vec{p}(t)=\Im \int_{\Bbb R^n}\bar{u}\nabla u dx,
\end{equation}
a quantity that has no definite sign. It turns out that one can also use this conservation law in the defocusing case and prove monotonicity formulas that are very useful in studying the global-in-time properties of the solutions at $t=\infty$. For most of these classical results the reader can consult
\cite{tc}, \cite{tt}.

The study of the problem at infinity is an attempt to describe and classify the asymptotic behavior-in-time
 for the global solutions. To handle this issue, one tries to compare the given nonlinear dynamics with suitably chosen simpler asymptotic dynamics. For the semilinear problem \eqref{nls}, the first obvious candidate
for the simplified asymptotic behavior is the free dynamics generated by the group $S(t)=e^{-it\Delta}$. The comparison
 between the two dynamics gives rise to the questions of the existence of wave operators and of the asymptotic completeness of the solutions. In the literature nowadays, the problem of asymptotic completeness is synonymous with the problem of scattering. More precisely, we have:
\\
{\it i)} Let $v_{+}(t)=S(t)u_{+}$ be the solution of the free equation. Does there exist a solution $u$ of equation \eqref{nls} which behaves asymptotically
 as $v_{+}$ as $t \rightarrow \infty$, typically in the sense that $\|u(t)-v_{+}\|_{H^{1}} \rightarrow 0, \ \ \mbox{as $t \rightarrow \infty$.}$ If this is
true, then one can define the map $\Omega_{+}: u_{+} \rightarrow u(0)$. The map is called the wave operator and
 the problem of existence of $u$ for given $u_{+}$ is referred to as the problem of the {\it{existence of the wave operator}}. The analogous problem
 arises as $t \rightarrow -\infty$.
\\
{\it ii)} Conversely, given a solution $u$ of \eqref{nls}, does there exist an asymptotic state $u_{+}$ such that $v_{+}(t)=S(t)u_{+}$ behaves
asymptotically as $u(t)$, in the above sense. If that is the case for any $u$ with initial data in $X$ for some $u_{+} \in X$,
 one says that {\it{asymptotic completeness}} holds in $X$.

In effect the existence of wave operators asks for the construction of global solutions that behave asymptotically as the solution of the free Schr\"odinger equation while the asymptotic completeness requires all solutions to behave asymptotically in this manner. It is thus not accidental that asymptotic completeness is a much harder problem than the existence of
the wave operators (except in the case of small data theory which
follows from the iterative methods of the local
 well-posedness theory).

 Asymptotic completeness for large data not only require a repulsive nonlinearity but also some decay for the nonlinear solutions. In the ideal scenario the nonlinear solution will obey the same decay properties of the linear Schr\"odinger equation. This decay of the linear problem can already establish weak quantum scattering in the energy space for example, but to say more, we usually need decay in an $L^p$ norm for the nonlinear solution. In this paper, we provide a summary of recent results that
demonstrates a straightforward method to obtain such estimates by taking advantage of the momentum conservation law \eqref{momentum}. A key example of these ideas is contained in the following generalized virial inequality of Lin and Strauss \cite{LinStrauss}. (We recall the proof of this inequality in the next section.)
\vskip 0.1 in
\begin{equation}\label{linstr}
\\
\int_{\Bbb R^n\times \R}(-\Delta \Delta a(x))|u(x,t)|^2dxdt+\lambda\int_{\Bbb R^n\times \R}(\Delta a(x)) {|u(x,t)|^{p+1}} dxdt \leq C
\end{equation}
\\
where $a(x)$ is a convex function, $u$ is a solution to \eqref{nls}, and $C$ a constant that depends only on the energy and mass bounds.
\vskip 0.1 in
An inequality of this form, which we will call a one-particle inequality, was first derived in the context of the
Klein-Gordon equation by Morawetz \cite{cm}, and then extended to the NLS equation in \cite{LinStrauss}. The inequality was applied to prove asymptotic completeness
 first for the nonlinear Klein-Gordon and then for the NLS equation in the papers by Morawetz and Strauss, \cite{ms}, and by Lin and Strauss, \cite{LinStrauss} for slightly more regular solutions in space dimension
$n \geq 3$. The case of general finite energy solutions for $n \geq 3$ was treated in \cite{gv} for the NLS and in \cite{gv1} for the Hartree equation. The treatment was then improved to the more difficult case of low dimensions by Nakanishi, \cite{kn}, \cite{kn1}.

 The bilinear a priori estimates that we outline here give stronger bounds on the solutions and in addition simplify the proofs of
 the results in the papers cited above. For a detailed summary of the method see \cite{gv2}. In the original paper by Morawetz, the weight function that was used was $a(x)=|x|$.
This choice has the advantage that the distribution $-\Delta \Delta (\frac{1}{|x|})$ is positive for $n \geq 3$. More precisely it is easy to compute that $\Delta a(x)=\frac{n-1}{|x|}$ and that
$$-\Delta \Delta a(x)=\left\{
\begin{array}{cc}
8\pi\delta(x), & \text{if} \quad  n=3\\
\frac{(n-1)(n-3)}{|x|^2}, & \text{if} \quad n \geq 4.
\end{array}
\right.
$$
In particular, the computation in \eqref{linstr} gives the following estimate for $n=3$ and $\lambda$ positive
\begin{equation}\label{linstr2}
\int_{\Bbb R}|u(t,0)|^2dt+\int_{\Bbb R^3\times \R}\frac{|u(x,t)|^{p+1}}{|x|} dxdt \leq C.
\end{equation}
Similar estimates are true in higher dimensions. The second, nonlinear term, or certain local versions of it, have played central role in the scattering theory for the
 nonlinear Schr\"odinger equation, \cite{jbou}, \cite{gv}, \cite{mg}, \cite{LinStrauss}.
The fact that in 3d, the bi-harmonic operator acting on the weight
$a(x)$ produces the $\delta-$measure can be exploited further. In \cite{ckstt4}, a quadratic Morawetz inequality was proved by correlating two nonlinear densities $\rho_1(x)$ and $\rho_2(y)$ and define as $a(x,y)$ the distance between $x$ and $y$ in 3d. The authors obtained an a priori estimate of the form $\int_{\Bbb R^3\times \R}|u(x,t)|^{4}dx \leq C$
for solutions that stay in the energy space.
A frequency localized version of this estimate has been
successfully implemented to remove the radial assumption of Bourgain, \cite{jbou}, and prove global well-posedness and scattering for the energy-critical (quintic) equation in 3d, \cite{ckstt5}. For $n\geq 4$ new quadratic Morawetz estimates were given in \cite{tvz1}. Finally in \cite{cgt} and in \cite{pv} these estimates were extended to all dimensions.

We should mention that taking as the weight function the distance between two points in $\R^n$ is not the only approach, see \cite{cgt1} for a recent example. Nowadays it is well understood that the bilinear Morawetz inequalities provide a unified approach for proving energy scattering for sub-critical solutions of the NLS when $p>1+\frac{4}{n}$ ($L^2$ super-critical nonlinearities). This last statement has been rigorously formalized only recently due to the work of the aforementioned authors, and a general exposition has been published in \cite{gv2}. Sub-energy solution scattering in the same range of powers has been initiated in \cite{ckstt4}. For the $L^2$-critical problem, scattering is a very hard problem, but it seems that the problem has now been resolved in a series of new papers by B. Dodson, \cite{dod,dod1,dod2}. For mass sub-critical solutions, scattering even in the energy space is a very hard problem, and is probably false. Nevertheless, two particle Morawetz estimates have been used for the problem of the existence (but not uniqueness) of the wave operator for mass subcritical problems, \cite{ht}. We have already mentioned their implementation to the hard problem of energy critical solutions in \cite{jbou}, \cite{mg}, and \cite{ckstt5}. Recent preprints have used these inequalities for the mass critical problem, \cite{dod}, and the energy super-critical problem, \cite{kv}. For a frequency localized one particle Morawetz inequality and its application to the scattering problem for the mass-critical equation with radial data see \cite{tvz2}.

\subsection{The Gross-Pitaevskii Hierarchy}

The $q$-GP hierarchy appears naturally in
the derivation, based on the BBGKY hierarchy of density matrices,
of the nonlinear Schr\"odinger equation
as the dynamical mean field limit of the manybody quantum dynamics of  Bose gases
with $(1+\frac q2)$-particle interactions.
Recently, this research area  has experienced some
remarkable progress, see \cite{esy1,esy2,ey,kiscst,klma,rosc, gmm-cmp, gmm-advm, xch-arma} and the references therein,
and also \cite{adgote,eesy,frgrsc,frknpi,frknsc,he,sp}.

Of particular importance for our work is the method developed in works of Erd\"os, Schlein, and Yau,
\cite{esy1,esy2,ey},
which consists of the following main steps:
\begin{enumerate}
\item One determines the BBGKY hierarchy of
marginal density matrices
for particle number $N$, and derives the Gross-Pitaevskii (GP)
hierarchy in the limit $N\rightarrow\infty$, in a scaling where
the particle interaction potential tends to a delta distribution; see also \cite{kiscst,sc}.

\item
One establishes uniqueness of solutions to the GP hierarchy.
This is the most difficult part of this analysis, and is obtained in \cite{esy1,esy2,ey}
by use of Feynman graph expansion methods inspired by quantum field theory.
It is subsequently verified that for factorized initial data
$$\gamma_0^{(k)}(\ux_k;\ux_k')=\prod_{j=1}^{k}\phi_0( x_{j})\overline{\phi_0( x_{j}^{\prime}})\,,$$
the solution of the GP hierarchy remains factorized  for all $t\in I\subseteq\R$,
$$\gamma^{(k)}(t,\ux_k;\ux_k')=\prod_{j=1}^{k}\phi(t,x_{j})\overline{\phi(t,x_{j}^{\prime})} \,,$$
if  $\phi(t)\in H^1(\R^d)$ solves the defocusing NLS,
$$
	i\partial_t\phi \, = \, - \Delta_x \phi \, + \, \lambda \, |\phi|^q\phi\,,
$$
for $t\in I\subseteq\R$, and $\phi(0)=\phi_0\in H^1(\R^d)$
with $q \in \{2,4\}$.  In other words, the solutions of the GP hierarchy
are governed by a cubic NLS for systems with 2-body interactions ($q=2$), \cite{esy1,esy2,ey,kiscst},
and quintic NLS for systems with 3-body interactions ($q=4$), \cite{chpa}.
\end{enumerate}

More recently, Klainerman and Machedon
have introduced an alternative method in \cite{klma} to prove the uniqueness of solutions to the
GP hierarchy for $q=2$ and $d=3$, in spaces defined via spacetime bounds on the density
matrices in the GP hierarchy; these spaces are different from the ones used in
\cite{esy1,esy2,ey}. The uniqueness result in \cite{klma} relies on the assumption of an a priori
spacetime bound on the density matrices.
For dimensions $d\leq2$, this assumption can be directly verified by exploiting energy conservation
in the BBGKY hierarchy in the limit $N\rightarrow\infty$, combined with a Sobolev type
inequality for density matrices.
This was recognized in
the paper \cite{kiscst} by Kirkpatrick, Schlein and Staffilani
where the authors prove uniqueness for the cubic case in $d=2$,
and establish the assumption made in \cite{klma}.
In \cite{chpa}, the corresponding problem in $d=1,2$ is solved for the quintic case.
Both  \cite{kiscst} and  \cite{chpa} involve a step where a certain spacetime norm is
controlled by using a Sobolev inequality for density matrices, and an a priori energy bound,
without exploiting the decay in time.
This approach is applicable for $d\leq2$, but not for dimension $d=3$.
In $d=3$,
a method is necessary that is truly based on spacetime norms; and such an approach
has recently been developed in \cite{chpa5}.

In \cite{chpa2},  Chen and Pavlovi\'{c} initiated
the study of the well-posedness of the Cauchy problem for
GP hierarchies with both focusing and defocusing interactions,
starting directly on the level of GP hierarchies, and independent of
the derivation from $N$-body Schr\"odinger dynamics.
Accordingly, the corresponding GP hierarchies are referred to as  {\em cubic} ($q=2$),
{\em quintic} ($q=4$), {\em focusing}, or {\em defocusing GP hierarchies},
depending on the type of the NLS governing the solutions obtained from
factorized initial conditions.
We note that for instance, it is currently not known how to rigorously derive a GP hierarchy from the
$N\rightarrow\infty$ limit of a BBGKY hierarchy with $L^2$-supercritical, attractive
interactions.

For the analysis of the Cauchy problem for $q$-GP hierarchies in \cite{chpa2},
the following Banach space of sequences
of $k$-particle marginal density matrices is introduced in \cite{chpa2}:
\eqn\label{bigG}
	\Gspace \, = \, \{ \, \Gamma \, = \, ( \, \gamma^{(k)}(x_1,\dots,x_k;x_1',\dots,x_k') \, )_{k\in\N}
	\, | \,
	\tr \gamma^{(k)} \, < \, \infty \, \} \,.
\eeqn
Given $\xi>0$,  the space
\eqn
	\cH_\xi^\alpha \, = \, \{\Gamma \, | \, \| \, \Gamma \, \|_{\cH_\xi^\alpha} \, < \, \infty \, \}
\eeqn
is endowed with the norm
\eqn\label{eq-KM-aprioriassumpt-0-1}
	\| \, \Gamma \, \|_{\cH_\xi^\alpha} \, := \,
	\sum_{k\in\N} \xi^k \, \| \, \gamma^{(k)} \, \|_{H^\alpha} \,,
\eeqn
where
\eqn\label{eq-gamma-norm-def-0-0}
	\| \gamma^{(k)} \|_{H^\alpha_k} & := &  \| S^{(k,\alpha)} \gamma^{(k)} \|_{L^2({\R}^{2kd})}
\eeqn
with\footnote{We note that the homogeneous version of the $H^{\alpha}_k$ norm, where 
$S^{(k, \alpha)}$ is replaced with 
$R^{(k,\alpha)}:=\prod_{j=1}^k |\nabla_{x_j}|^\alpha |\nabla_{x_j'}|^\alpha$
is the norm considered in \cite{klma}.}  
$$
	S^{(k,\alpha)}:=\prod_{j=1}^k\bra\nabla_{x_j}\ket^\alpha\bra\nabla_{x_j'}\ket^\alpha.
$$
If $\Gamma\in\cH_\xi^\alpha$, then $\xi^{-1}$ an upper bound on the typical $H^\alpha$-energy per particle.

In \cite{chpa2},  the existence and uniqueness of solutions
for energy subcritical focusing and defocusing  cubic
and quintic GP hierarchies is proven in a subspace of $\cH_\xi^\alpha$.
The parameter $\alpha$ determines the regularity of the solution,
and results of \cite{chpa2} hold for  $\alpha\in\alphaset(d,q)$ where
\eqn\label{eq-alphaset-def-0}
    \alphaset(d,q) \, := \, \left\{
    \begin{array}{cc}
    (\frac12,\infty) & {\rm if} \; d=1 \\
    (\frac d2-\frac{1}{2(q-1)}, \infty) & {\rm if} \; d\geq2 \; {\rm and} \; (d,q)\neq(3,2)\\
    \big[1,\infty) & {\rm if} \; (d,q)=(3,2) \,,
    \end{array}
    \right.
\eeqn
in dimensions $d\geq1$, and where $q=2$ for the cubic, and $q=4$ for the quintic GP hierarchy\footnote{
The condition on $\alpha$ comes as a consequence of a Strichartz type estimate 
which was first obtained for $q=2$ in \cite{klma} and for $q=4$ in \cite{chpa}. The version 
used in \cite{chpa2} was formulated in Proposition 1 in \cite{chpa2}. We also recall that the scaling invariant 
version of a Strichartz estimate for homogeneous norms when $q=2$ was obtained in \cite{xch-arx}.}.  
The proof involves a Picard fixed point argument, and holds for various dimensions $d$,
without any requirement on factorization.
The solutions derived in \cite{chpa2}  satisfy a space-time bound corresponding to the
one used (as an a priori assumption) in the work of Klainerman and Machedon, \cite{klma}.
The parameter $\xi>0$ is determined by the initial condition, and it sets the energy scale of
the given Cauchy problem.

The work \cite{chpa2} motivated further study of the Cauchy problem for q-GP hierarchy:
\begin{itemize}
\item In \cite{chpatz1},  we identified a conserved energy functional
$E_1(\Gamma(t)) = E_1(\Gamma(0))$
describing the average energy per particle, and we proved
virial identities for solutions of GP hierarchies.
In particular, we use these ingredients to prove that for $L^2$-critical
and supercritical focusing GP hierarchies, blowup occurs
whenever $E_1(\Gamma_0)<0$, and when the variance is finite.
We note that prior to  \cite{chpatz1}, no exact conserved energy functional on
the level of the GP hierarchy was identified in any of the previous works,
including \cite{kiscst} and \cite{esy1,esy2}.

\item In \cite{chpa3}, new higher order energy functionals were introduced, and
proven to be conserved quantitites for energy subcritical defocusing, and
$L^2$ subcritical (de)focusing GP hierarchies, in spaces similar to those
used by Erd\"os, Schlein and Yau in  \cite{esy1,esy2}.
By use of this tool, a priori $H^1$ bounds for positive semidefinite solutions
are proven in the same spaces.
Moreover, global well-posedness was obtained for positive semidefinite solutions
in the spaces studied in the works of Klainerman and Machedon, \cite{klma}, and in \cite{chpa2}.

\item A new proof of existence of solutions  to  cubic and quintic GP hierarchies
for focusing and defocusing interactions was obtained in \cite{chpa4}.
It does
not require the a priori bound on the spacetime norms, which was introduced in the work of
Klainerman and Machedon, \cite{klma}, and used in the earlier work of Chen and Pavlovi\'{c}
\cite{chpa2}.

\item Very recently,  a new derivation of the defocusing cubic GP hierarchy has been obtained
in \cite{chpa5} for dimensions $d=2,3$, which crucially involves
generalizations of the tools developed in the above
mentioned paper \cite{chpa4}.  In particular, it is established in
\cite{chpa5} that the GP hierarchy obtained from the $N\rightarrow\infty$ limit
of the corresponding BBGKY hierarchy, is contained in the space introduced by
Klainerman and Machedon in \cite{klma} based on their spacetime norms.
While these results do not assume factorization of the solutions,
consideration of the special case of factorized solutions yields
a new derivation of the cubic, defocusing NLS in $d=2,3$.
\end{itemize}

In this paper, we extend the study of Morawetz type identities for the GP hierarchy,
inspired by our proof of a virial identity for GP hierarchies in \cite{chpatz1}.
More precisely, we derive an interaction Morawetz identity,
and as a consequence, a regular one-particle Morawetz identity for
solutions of the GP hierarchy.
These calculations are carried out in Section \ref{sec-intMorawetz-1}
of the paper.


\vskip 0.1 in
\section{The nonlinear Schr\"odinger equation.}
We start with the equation
\begin{equation}\label{nls2}
iu_{t}+\Delta u=\lambda |u|^{p-1}u
\end{equation}
with $p \geq 1$ and $\lambda \in \R$. We use Einstein's summation convention throughout this section. According to this convention, when an index variable appears twice in a single term, once in an upper (superscript) and once in a lower (subscript) position, it implies that we are summing over all of its possible values. We will also write $\nabla_{j}u$ for $\frac{\partial u}{\partial x_{j}}$. For a function $a(x,y)$ defined on $\R^n \times \R^n$ we define
$\nabla_{x,j}\ a(x,y)=\frac{\partial a(x,y)}{\partial x_j}$ and similarly for $\nabla_{x,k}\ a(x,y)$.
\\
\\
We define the mass density $\rho$ and the momentum vector $\vec{p}$, by the relations
$$\rho=|u|^2,\ \ \ \ \ \ \ \ \ \ \ \ p_{k}=\Im (\bar{u}\nabla_{k}u).$$
It is well known, \cite{tc}, that smooth solutions to the semilinear Schr\"odinger equation satisfy mass and momentum conservation. The local conservation of mass reads
\begin{equation}\label{conmass}
\partial_{t}\rho+2div{\vec{p}}=\partial_{t}\rho+2\nabla_{j}p^{j}=0
\end{equation}
and the local momentum conservation is
\begin{equation}\label{conmom}
\partial_{t}p^{j}+\nabla^{k}\Big(\delta_{k}^{j}\big(-\frac{1}{2}\Delta \rho +\lambda \frac{p-1}{p+1}|u|^{p+1}\big)+\sigma_{k}^{j} \Big)=0
\end{equation}
where the symmetric tensor $\sigma_{jk}$ is given by
$$\sigma_{jk}=2\Re(\nabla_{j}u\nabla_{k}\overline{u}).$$
Notice that the term $\lambda\frac{p-1}{p+1}|u|^{p+1}$ is the only nonlinear term that
appears in the expression. One can express the local conservation laws purely in terms of the mass density $\rho$ and the momentum $\vec{p}$ if we write
$$\lambda \frac{p-1}{p+1}|u|^{p+1}=2^{\frac{p+1}{2}}\lambda \frac{p-1}{p+1}\rho ^{\frac{p+1}{2}}$$
and
$$\sigma_{jk}=2\Re(\nabla_{j}u\nabla_{k}\overline{u})=\frac{1}{\rho}(2p_{j}p_{k}+\frac{1}{2}\nabla_{j}\rho \nabla_{k}\rho),$$
but we won't use this formulation in this paper. We are ready to state the main theorem of this section:
\begin{theorem}\label{NLSthm}
\cite{cgt, ckstt4, pv, tvz1} Consider $u\in C_{t}(\R;C_{0}^{\infty}(\R^n))$ a smooth and compactly supported solution to \eqref{nls2} with $u(x,0)=u(x)\in C_{0}^{\infty}(\R^n)$. Then for $n \geq 2$
 we have that
 $$C\|D^{-\frac{n-3}{2}}(|u|^2)\|_{L_{t}^{2}L_{x}^{2}}^{2}+(n-1)\lambda \frac{p-1}{p+1}\int_{\R_{t}}\int_{\R_{x}^n \times \R_{y}^n}\frac{|u(y,t)|^2|u(x,t)|^{p+1}}{|x-y|}dxdydt$$
$$\leq \|u_{0}\|_{L^2}^{2}\sup_{t\in \R}|M_{y}(t)|,$$
where $$M_{y}(t)=\int_{\R^n}\frac{x-y}{|x-y|} \cdot \Im \big( \overline{u}(x)\nabla u(x)\big)dx,$$ $D^{\alpha}$ is defined on the Fourier side as $\widehat{D^{\alpha}f}(\xi)=|\xi|^{\alpha}\widehat{u}(\xi)$ for any $\alpha \in \R$ and $C$ is a positive constant that depends only on $n$, \cite{es}. For $n=1$ the estimate is
$$\|\partial_{x}(|u|^2)\|_{L_{t}^{2}L_{x}^{2}}^{2}+\lambda \frac{p-1}{p+1}\|u\|_{L_{t}^{p+3}L_{x}^{p+3}}^{p+3}\leq \frac{1}{2}\|u_{0}\|_{L^2}^{3}\sup_{t\in \R}\|\partial_{x}u\|_{L^2}.$$
\end{theorem}

$\;$

{\bf Remarks on Theorem \ref{NLSthm}.}
\vskip 0.2 in
\noindent
1. By the Cauchy-Schwarz inequality it follows that for any $n \geq 2$,
$$\sup_{0,t}|M_{y}(t)|\lesssim \|u_{0}\|_{L^2}\sup_{t \in \R}\|\nabla u(t)\|_{L^2}.$$
A variant of Hardy's inequality gives
$$\sup_{0,t}|M_{y}(t)|\lesssim \sup_{t \in \R}\|u(t)\|_{\dot {H}^{\frac{1}{2}}}^2,$$
For details, see \cite{gv2}.
\vskip 0.1 in
\noindent
2. Concerning our main theorem, we note that both the integrated functions in the second term on the left hand side of the inequalities are positive. Thus when $\lambda>0$, which corresponds to the defocusing case, and for $H^1$ data say, we obtain for $n\geq 2$:
$$\|D^{-\frac{n-3}{2}}(|u|^2)\|_{L_{t}^{2}L_{x}^{2}}\lesssim \|u_{0}\|_{L^2}^{\frac{3}{2}}\sup_{t \in \R}\|\nabla u(t)\|_{L^2}^{\frac{1}{2}}\lesssim M(u_{0})^{\frac{3}{2}}E(u_{0})^{\frac{1}{2}},$$
and for $n=1$
$$\|\partial_{x}(|u|^2)\|_{L_{t}^{2}L_{x}^{2}}^{2}\lesssim \|u_{0}\|_{L^2}^{\frac{3}{2}}\sup_{t \in \R}\|\partial_{x} u(t)\|_{L^2}^{\frac{1}{2}}\lesssim M(u_{0})^{\frac{3}{2}}E(u_{0})^{\frac{1}{2}}.$$
These are easy consequences of the conservation laws of mass \eqref{mass} and energy \eqref{energy}. They provide the global a priori estimates that are used in quantum scattering in the energy space, \cite{gv2}.
\vskip 0.1 in
\noindent
3. Analogous estimates hold for the case of the Hartree equation $iu_{t}+\Delta u=\lambda (|x|^{-\gamma}\star\ |u|^2)u$ when $0<\gamma <n$, $n\geq 2$. For the details, see \cite{ht}. We should point out that for $0<\gamma \leq 1$ scattering fails for the Hartree equation, \cite{ht1}, and thus the estimates given in \cite{ht} for $n \geq 2$ cover all the interesting cases. We also expect the Hartree quadratic Morawetz estimates that were established in \cite{ht} to extend to the case of the
Hartree hierarchy, for which factorized solutions are determined by the Hartree equation. A detailed analysis of this problem for GP hierarchies is presented in section \ref{sec-intMorawetz-1}, below.
\vskip 0.1 in
\noindent
4. Take $\lambda >0$. The expression $$\|D^{-\frac{n-3}{2}}(|u|^2)\|_{L_{t}^{2}L_{x}^{2}},$$ for $n=3$, provides an estimate for the $L_{t}^{4}L_{x}^{4}$ norm of the solution.
For $n=2$ by Sobolev embedding one has that
$$\|u\|_{L_{t}^{4}L_{x}^{8}}^2=\||u|^2\|_{L_{t}^{2}L_{x}^{4}}\lesssim \|D^{\frac{1}{2}}\big(|u|^2\big)\|_{L_{t}^{2}L_{x}^{2}} \lesssim C_{M(u_{0}),E(u_{0})}.$$
For $n\geq 4$ the power of the $D$ operator is negative but some harmonic analysis and interpolation with the trivial inequality
$$\|D^{\frac{1}{2}}u\|_{L_{t}^{\infty}L_{x}^{2}}\lesssim \|u\|_{L_{t}^{\infty}\dot{H}_{x}^{\frac{1}{2}}}$$
provides an estimate in a Strichartz norm. For the details see \cite{tvz1}.
\vskip 0.1 in
\noindent
5. In the defocusing case all the estimates above give a priori information for the $\dot{H}^{\frac{1}{4}}$-critical Strichartz norm. We remind the reader that the $\dot{H}^{s}$-critical Strichartz norm is $\|u\|_{L_{t}^{q}L_{x}^{r}}$ where the pair $(q,r)$ satisfies $\frac{2}{q}+\frac{n}{r}=\frac{n}{2}-s$. In principle the correlation of $k$ particles will provide a priori information for the $\dot{H}^{\frac{1}{2k}}$ critical Strichartz norm. In 1d an estimate that provides a bound on the $\dot{H}^{\frac{1}{8}}$ critical Strichartz norm has been given in \cite{chvz}.
\vskip 0.1 in
\noindent
6. To make our presentation easier we considered smooth solutions of the NLS equation. To obtain the estimates in Theorem \ref{NLSthm} for arbitrary $H^1$ functions we have to regularize the solutions and then take a limit. The process is described in \cite{gv2}.
\vskip 0.1 in
\noindent
7. A more general bilinear estimate can be proved if one correlates two different solutions (thus considering different density functions $\rho_1$ and $\rho_2$). Unfortunately, one can obtain useful estimates only for $n \geq 3$. The proof is based on the fact that $-\Delta^2|x|$ is a positive distribution only for $n \geq 3$. For details the reader can check \cite{ckstt4}. Our proof shows that the diagonal case when $\rho_1=\rho_2=|u|^2$
provides useful monotonicity formulas in all dimensions.
\begin{proof}
We define the Morawetz action centered at zero by
\eqn\label{eq-NLS-simpMorawetz}
	M_{0}(t)=\int_{\R} \nabla a(x) \cdot \vec{p}(x)\ dx,
\eeqn
where the weight function $a(x):\R^n \to \R$ is for the moment arbitrary. The minimal requirements on $a(x)$ call for the matrix of the second partial derivatives $\partial_{j}\partial_{k}a(x)$ to be positive definite. Throughout our paper we will take $a(x)=|x|$, but many estimates can be given with different weight functions, see for example \cite{cgt1} and \cite{kv}. If we differentiate the Morawetz action with respect to time we obtain:
$$\partial_{t}M_{0}(t)=\int_{\R^n}\nabla a(x)\cdot \partial_{t} \vec{p}(x)\ dx=\int_{\R^n}\pj a(x) \partial_{t} p^{j}(x)\ dx$$
$$=\int_{\R^n}\big(\pj \pkup a(x)\big)\delta_{k}^{j}\big(-\frac{1}{2}\Delta \rho +\lambda\frac{p-1}{p+1}|u|^{p+1}\big)dx+2\int_{\R^n}\big(\pj \pkup a(x)\big)\Re \big(\pjup\overline{u}\pk u\big)dx,$$
where we use equation \eqref{conmom}. We rewrite and name the equation as follows
\begin{equation}\label{doublestar}
\partial_{t}M_{0}(t)=\int_{\R^n}\Delta a(x)\big(-\frac{1}{2}\Delta \rho +\lambda\frac{p-1}{p+1}|u|^{p+1}\big)dx+2\int_{\R^n}\big(\pj \pkup a(x)\big)\Re \big(\pjup\overline{u}\pk u\big)dx.
\end{equation}
Notice that for $a(x)=|x|$ the matrix $\pj \pk a(x)$ is positive definite and the same is true if we translate the weight function by any point $y \in \R^n$ and consider $\nabla_{x,j}\nabla^{x,k}a(x-y)$ for example. That is for any vector function on $\R^n$, $\{v_{j}(x)\}_{j=1}^{n}$, with values on $\R$ or $\C$ we have that
$$\int_{\R^n}\big(\pj \pkup a(x)\big)v^{j}(x)v_{k}(x)dx \geq 0.$$
To see this, observe that for $n \geq 2$ we have $\pj a=\frac{x_j}{|x|}$ and $\pj \pk a=\frac{1}{|x|}\big(\delta_{kj}-\frac{x_jx_k}{|x|^2}\big)$. Summing over $j=k$ we obtain $\Delta a(x)=\frac{n-1}{|x|}$. Then
$$\pj \pkup a(x)v^{j}(x)v_{k}(x)=\frac{1}{|x|}\big(\delta_{kj}-\frac{x_jx_k}{|x|^2}\big)v^{j}(x)v_{k}(x)=\frac{1}{|x|}\Big(|\vec{v}(x)|^2-\big( \frac{x\cdot \vec{v}(x)}{|x|}\big)^2 \Big)\geq 0$$
by the Cauchy-Schwarz inequality. Notice that it does not matter if the vector function is real or complex valued for this inequality to be true. In dimension one \eqref{doublestar} simplifies to
\begin{equation}\label{doublestar1}
\partial_{t}M_{0}(t)=\int_{\R}a_{xx}(x)\big(-\frac{1}{2}\Delta \rho +\lambda\frac{p-1}{p+1}|u|^{p+1}+2|u_{x}|^{2}\big)dx.
\end{equation}
In this case for $a(x)=|x|$, we have that $a_{xx}(x)=2\delta(x)$.
Since the identity \eqref{doublestar} does not change if we translate the weight function by $y\in \R^n$ we can define the Morawetz action with center at $y\in \R^n$ by
$$M_{y}(t)=\int_{\R^n} \nabla a(x-y) \cdot \vec{p}(x)\ dx.$$
We can then obtain like before
\begin{align*}\label{doublestar2}
\partial_{t}M_{y}(t)=\int_{\R^n}\Delta_{x} a(x-y)\big(-\frac{1}{2}\Delta \rho+\lambda\frac{p-1}{p+1}|u|^{p+1}\big)dx\\
+2\int_{\R^n}\big(\nabla_{x,j} \nabla^{x,k} a(x-y)\big)\Re \big(\nabla^{x,j}\overline{u}\nabla_{x,k} u\big)dx.
\end{align*}
We now define the two-particle Morawetz action
$$M(t)=\int_{\R_{y}^{n}} |u(y)|^2 M_{y}(t)\ dx$$ and differentiate with respect to time. Using the identity above and the local conservation of mass law we obtain four terms
$$\partial_{t}M(t)=\int_{\R_{y}^{n}} |u(y)|^2 \partial_{t} M_{y}(t)\ dx+\int_{\R_{y}^{n}} \partial_{t}\rho(y) M_{y}(t)\ dx$$
$$=\int_{\R_{y}^n \times \R_{x}^{n}}|u(y)|^2\Delta_{x} a(x-y)\big(-\frac{1}{2}\Delta \rho+\lambda\frac{p-1}{p+1}|u|^{p+1}\big)dxdy$$
$$+2\int_{\R_{y}^n\times \R_{x}^{n}}|u(y)|^2\big(\nabla_{x,j} \nabla^{x,k} a(x-y)\big)\Re \big(\nabla^{x,j}\overline{u}\nabla_{x,k} u\big)dxdy$$
$$-2\int_{\R_{y}^n\times \R_{x}^{n}}\nabla^{y,j}p_{j}(y)\nabla_{x,k}a(x-y)p^{k}(x)dxdy$$
$$=I+II+III+2\int_{\R_{y}^n\times \R_{x}^{n}}p_{j}(y)\nabla^{y,j}\nabla_{x,k}a(x-y)p^{k}(x)dxdy$$
by integration by parts with respect to the $y-$variable. Since
$$\nabla^{y,j}\nabla_{x,k}a(x-y)=-\nabla^{x,j}\nabla_{x,k}a(x-y)$$
we obtain that
\begin{equation}\label{mainestimate}
\partial_{t}M(t)=I+II+III-2\int_{\R_{y}^n\times \R_{x}^{n}}\nabla^{x,j}\nabla_{x,k}a(x-y)p_{j}(y)p^{k}(x)dxdy
\end{equation}
$$=I+II+III+IV$$
where
$$I=\int_{\R_{y}^n \times \R_{x}^{n}}|u(y)|^2\Delta_{x} a(x-y)\big(-\frac{1}{2}\Delta \rho \big)dxdy,$$
$$II=\int_{\R_{y}^n \times \R_{x}^{n}}|u(y)|^2\Delta_{x} a(x-y)\big(\lambda\frac{p-1}{p+1}|u|^{p+1}\big)dxdy,$$
$$III=2\int_{\R_{y}^n\times \R_{x}^{n}}|u(y)|^2\big(\nabla_{x,j} \nabla^{x,k} a(x-y)\big)\Re \big(\nabla^{x,j}\overline{u}\nabla_{x,k} u\big)dxdy,$$
$$IV=-2\int_{\R_{y}^n\times \R_{x}^{n}}\nabla^{x,j}\nabla_{x,k}a(x-y)p_{j}(y)p^{k}(x)dxdy.$$
\\
{\bf Claim:} $III+IV\geq 0$. Assume the claim. Since $\Delta_{x}a(x-y)=\frac{n-1}{|x-y|}$ we have that
$$\partial_{t}M(t) \geq \frac{n-1}{2}\int_{\R_{y}^n \times \R_{x}^{n}}\frac{|u(y)|^2}{|x-y|}\big(-\Delta \rho \big)dxdy+
(n-1)\lambda\frac{p-1}{p+1}\int_{\R_{y}^n \times \R_{x}^{n}}\frac{|u(y)|^2}{|x-y|}|u(x)|^{p+1}dxdy.$$
But recall that on one hand we have that $-\Delta=D^2$ and on the other that the distributional Fourier transform of $\frac{1}{|x|}$ for any $n \geq 2$ is $\frac{c}{|\xi|^{n-1}}$ where $c$ is a positive constant depending only on $n$. Thus we can define
$$D^{-(n-1)}f(x)=c\int_{\R^n}\frac{f(y)}{|x-y|}dy$$
and express the first term as
$$\frac{n-1}{2}\int_{\R_{y}^n \times \R_{x}^{n}}\frac{|u(y)|^2}{|x-y|}\big(-\Delta \rho \big)dxdy=c\frac{n-1}{2}<D^{-(n-1)}|u|^2,\ D^2|u|^2>=
C\|D^{-\frac{n-3}{2}}|u|^2\|_{L_{x}^{2}}^{2}$$
by the usual properties of the Fourier transform for positive and real functions. Integrating from $0$ to $t$ we obtain the theorem in the case that $n \geq 2$.
\\
\\
{\bf Proof of the claim:} Notice that
$$III+IV=2\int_{\R_{y}^n\times \R_{x}^{n}}\nabla_{x,j} \nabla^{x,k} a(x-y)\Big(|u(y)|^2\Re \big(\nabla^{x,j}\overline{u}(x)\nabla_{x,k} u(x)\big)-p^{j}(y)p_{k}(x)\Big)dxdy $$
$$=2\int_{\R_{y}^n\times \R_{x}^{n}}\nabla_{x,j} \nabla^{x,k} a(x-y)\Big(\frac{\rho(y)}{\rho(x)}\Re \big(u(x)(\nabla^{x,j}\overline{u}(x)) \overline{u}(x)(\nabla_{x,k}u(x))\big)-p^{j}(y)p_{k}(x)\Big)dxdy.$$
Since $$\nabla_{x,j} \nabla_{x,k} a(x-y)=\nabla_{y,j} \nabla_{y,k} a(y-x)$$
by exchanging the roles of $x$ and $y$ we obtain the same inequality and thus
$$III+IV=\int_{\R_{y}^n\times \R_{x}^{n}}\nabla_{x,j} \nabla^{x,k} a(x-y)\Big(\frac{\rho(y)}{\rho(x)}\Re \big(u(x)(\nabla^{x,j}\overline{u}(x))\overline{u}(x)(\nabla_{x,k}u(x))\big)-p^{j}(y)p_{k}(x)$$
$$+\frac{\rho(x)}{\rho(y)}\Re \big(u(y)(\nabla^{y,j}\overline{u}(y))\overline{u}(y)(\nabla_{y,k}u(y))\big)-p^{j}(x)p_{k}(y)\Big)dxdy.$$
Now set $z_1=\overline{u}(x)\nabla_{x,k}u(x)$ and $z_2=\overline{u}(x)\nabla^{x,j}u(x)$ and apply the identity
$$\Re(z_1\bar{z}_2)=\Re(z_1)\Re(z_2)+\Im(z_1)\Im(z_2)$$ to obtain
$$\Re \big(u(x)(\nabla^{x,j}\overline{u}(x))\overline{u}(x)(\nabla_{x,k}u(x))\big)=\Re\big(\overline{u}(x)\nabla_{x,k}u(x)\big)\Re\big(\overline{u}(x)\nabla^{x,j}u(x)\big)$$
$$+\Im\big(\overline{u}(x)\nabla_{x,k}u(x)\big)\Im\big(\overline{u}(x)\nabla^{x,j}u(x)\big)=\frac{1}{4}\nabla_{x,k}\rho(x)\nabla^{x,j}\rho(x)+p_{k}(x)p^{j}(x)$$
and similarly
$$\Re \big(u(y)(\nabla^{y,j}\overline{u}(y))\overline{u}(y)(\nabla_{y,k}u(y))\big)=\frac{1}{4}\nabla_{y,k}\rho(y)\nabla^{y,j}\rho(y)+p_{k}(y)p^{j}(y).$$
Thus
$$III+IV=\frac{1}{4}\int_{\R_{y}^n\times \R_{x}^{n}}\nabla_{x,j} \nabla^{x,k} a(x-y)\frac{\rho(y)}{\rho(x)}\nabla_{x,k}\rho(x)\nabla^{x,j}\rho(x)dxdy$$
$$+\frac{1}{4}\int_{\R_{y}^n\times \R_{x}^{n}}\nabla_{y,j} \nabla^{y,k} a(x-y)\frac{\rho(x)}{\rho(y)}\nabla_{y,k}\rho(y)\nabla^{y,j}\rho(y)dxdy$$
$$+\int_{\R_{y}^n\times \R_{x}^{n}}\nabla_{y,j} \nabla^{y,k} a(x-y)\Big(\frac{\rho(y)}{\rho(x)}p_k(x)p^j(x)+\frac{\rho(x)}{\rho(y)}p_k(y)p^j(y)-p_k(x)p^j(y)-p_k(y)p^j(x)\Big)dxdy.$$
Since the matrix $\nabla_{x,j} \nabla^{x,k} a(x-y)=\nabla_{y,j} \nabla^{y,k} a(x-y)$ is positive definite, the first two integrals are positive. Thus,
$$III+IV\geq $$
$$\int_{\R_{y}^n\times \R_{x}^{n}}\nabla_{x,j} \nabla^{x,k} a(x-y)\Big(\frac{\rho(y)}{\rho(x)}p_k(x)p^j(x)+\frac{\rho(x)}{\rho(y)}p_k(y)p^j(y)-p_k(x)p^j(y)-p_k(y)p^j(x)\Big)dxdy.$$
Now if we define the two point vector
$$J_{k}(x,y)=\sqrt{\frac{\rho(y)}{\rho(x)}}p_{k}(x)-\sqrt{\frac{\rho(x)}{\rho(y)}}p_{k}(y)$$
we obtain that
$$
	III+IV\geq \int_{\R_{y}^n\times \R_{x}^{n}}\nabla_{x,j} \nabla^{x,k} a(x-y)J^{j}(x,y)
	J_{k}(x,y)dxdy\geq 0 \,
$$
and we are done.

The proof when $n=1$ is easier. First, an easy computation shows that if $a(x,y)=|x-y|$ then $\partial_{xx}a(x,y)=2\delta(x-y)$. In this case from \eqref{mainestimate} we obtain
$$\partial_{t}M(t)=\int_{\R_{y} \times \R_{x}}|u(y)|^{2}2\delta(x-y)\big(-\frac{1}{2}\rho_{xx} \big)dxdy+2\int_{\R}|u(x)|^2\big(\lambda\frac{p-1}{p+1}|u(x)|^{p+1}\big)dx$$
$$+4\int_{\R}|u(x)|^2|u_{x}|^2dx-4\int_{\R}p^2(x)dx.$$
But
$$\int_{\R_{y} \times \R_{x}}|u(y)|^{2}2\delta(x-y)\big(-\frac{1}{2}\rho_{xx} \big)dxdy=\int_{\R}\Big(\partial_{x}|u(x)|^2\Big)^2dx.$$
In addition a simple calculation shows that
$$|u(x)|^2|u_{x}|^2=\Big(\Re(\overline{u}u_{x})\Big)^2+\Big(\Im(\overline{u}u_{x})\Big)^2=\frac{1}{4}\Big(\partial_{x}|u|^2\Big)^2+p^2(x).$$
Thus
$$4|u(x)|^2|u_{x}|^2-4p^2(x)=\Big(\partial_{x}|u|^2\Big)^2$$
and the identity becomes
\eqn\label{eq-dtM-NLS-1}
	\partial_{t}M(t)=2\int_{\R}\Big(\partial_{x}|u|^2\Big)^2dx+2\int_{\R}
	|u(x)|^2\big(\lambda\frac{p-1}{p+1}|u(x)|^{p+1}\big)dx
\eeqn
which finishes the proof of the theorem.
\end{proof}

\section{ Morawetz identities for the GP hierarchy}
\label{sec-intMorawetz-1}

In this section, we derive one-particle Morawetz inequalities for GP hierarchies in
Theorem \ref{thm-Morawetz-main-2},
and interaction Morawetz identities for   GP hierarchies
in Theorem \ref{thm-Morawetz-main-1},
below, for $1\leq n\leq3$ dimensions.
For simplicity of exposition, we will only present the case of {\em cubic} GP hierarchies
here, that is, $q$-GP hierarchies with $q=2$.
The case of quintic or higher degree $q$-GP hierarchies ($q\geq4$, $q\in 2\N$) can be
treated in a completely analogous manner.

For convenience of exposition,  we assume that $\Gamma=(\gamma^{(k)})$ solves
the cubic GP hierarchy \eqref{intro-GP} in $\cH_\xi^2$ (the corresponding local well-posedness
theory is covered in
\cite{chpa2}), with  interaction terms
defined in \eqref{eq-def-b} with $q=2$.
\footnote{As a result, all expressions in the steps below can easily seen to be well-defined.
By adopting the arguments of \cite{chpa3}, it in fact suffices to consider solutions
in $\cH_\xi^1$.}

\subsection{One-particle Morawetz identities for the cubic GP hierarchy}

The density function corresponding to the one-particle marginal is defined by
\eqn
	\rho(x) & := & \gamma^{(1)}(x;x) \,.
\eeqn
One can straightforwardly verify that
\eqn
	\partial_t\rho(x) & = &
	\int du \, du' \, e^{i(u-u')x} \, \partial_t\widehat\gamma^{(1)}(u;u')
	\nonumber\\
	& = &
	\frac1i\int du \, du' \, e^{i(u-u')x}(u^2 - (u')^2) \, \widehat\gamma^{(1)}(u;u')
	\nonumber\\
	&&\hspace{2cm}
	+ \, \frac\lambda i\int du \, du' \, e^{i(u-u')x}\, \widehat{B_{1,2}\gamma^{(2)}}(u;u')
	\label{eq-rhoderiv-aux-1}\\
	& = &
	\frac1i\int du \, du' \, e^{i(u-u')x}(u+u')(u-u') \, \widehat\gamma^{(1)}(u;u')
	\nonumber\\
	& = &
	- \, \nabla_x \cdot \int du \, du' \, e^{i(u-u')x}(u+u') \, \widehat\gamma^{(1)}(u;u') \,,
\eeqn
so that in analogy with \eqref{conmass},
\eqn
	\partial_t\rho(x) \, + \, 2 \nabla_x\cdot P(x) \, = \, 0 \,,
\eeqn
with
\eqn\label{eq-Pmom-def-1}
	P(x) & := &\,
	 \int du \, du' \,  e^{i(u-u')x } \, \frac{u+u'}2 \, \widehat\gamma^{(1)}(u;u')
\eeqn
is the momentum operator,
see also \cite{chpatz1}.
The fact that the interaction term \eqref{eq-rhoderiv-aux-1} equals zero is proven in
eqs. (5.5) - (5.8) in \cite{chpatz1}.
We are here adopting conventions analogous to those
in the previous chapter applied to the NLS.
In the sequel, we will suppress the dependence on $t$ from the notation, for simplicity.

We define the Morawetz action
\eqn
	M_a \, := \, \int dx \, \nabla a(x) \cdot P(x) \,
\eeqn
in analogy to \eqref{eq-NLS-simpMorawetz}.
The time derivative is given by
\eqn
	\partial_t M_a \, = \, \int dx \, \nabla a(x) \cdot \partial_t P(x)  \,.
\eeqn
Then, we obtain the following version of
the regular Morawetz identity.

\begin{theorem}
\label{thm-Morawetz-main-2}
Under the conditions formulated above, the {\em one-particle Morawetz identity}
\eqn
	\partial_t M_a
	&=&-\frac{1}{2} \, \int dx \,
	(\Delta_x \Delta_x a(x))   \, \gamma^{(1)}(x;x)
	\nonumber\\
	&&+ \, \frac\lambda2 \int dx \, (\Delta_x a(x))
	\gamma^{(2)}(x,x;x,x)
	\nonumber\\
	&&+ 2 \, \Re \int dx dx' \delta(x-x') \,
	\sum_{j,\ell} \, \big( \, \partial_{x_j}  \partial_{x_\ell} a(x) \, \big)
	\partial_{x_\ell}\partial_{x_j'}
	\gamma^{(1)}(x;x')   \,.
	\label{eq-simpMorawetz-final-1}
\eeqn
holds for solutions of the cubic GP hierarchy.
\end{theorem}

We defer the proof to section \ref{ssec-simpMorawetz-1}.

\subsubsection{Factorized solutions}
Substituting factorized solutions of the form
\eqn\label{eq-gammak-factorized-1}
	\gamma^{(k)}(t,\ux_k,\ux_k')=\prod_{j=1}^k\phi(t,x_j)\overline{\phi(t,x_j')} \,,
\eeqn
where
\eqn
	i\partial_t\phi(t,x) \, + \, \Delta_x \phi(t,x) \, = \,
	\lambda \, |\phi(t,x)|^2 \, \phi(t,x)
\eeqn
with initial data $\phi(0,\,\cdot\,)=\phi_0\in H^1$,
we obtain the following result.
\eqn
	\partial_t M_a
	&=&-\frac{1}{2} \, \int dx \,
	(\Delta_x \Delta_x a(x))   \, |\phi(t,x)|^2
	\nonumber\\
	&&+ \, \frac\lambda2 \int dx \, (\Delta_x a(x))
	|\phi(t,x)|^4
	\nonumber\\
	&&+  2 \, \Re \int dx   \,
	\sum_{j,\ell} \, \big( \, \partial_{x_j}  \partial_{x_\ell} a(x) \, \big)
	(\partial_{x_\ell}\phi(t,x)) \, \overline{\partial_{x_j'}  \phi(t,x')}   \,.
\eeqn
This corresponds to the one-particle Morawetz identity \eqref{doublestar} for the
NLS, in the cubic case $p=3$.

\subsection{Interaction Morawetz identities for the cubic GP hierarchy}

In this section, we derive  interaction   Morawetz identities for
GP hierarchies which generalize those for the NLS.

\subsubsection{Morawetz action}

We write
\eqn
	\gamma^{(2)}(x,y;x',y') \, = \, \int du \, du' \,  dv \, dv' \,   e^{iux-iu'x'} e^{ivy-iv'y'}
	\widehat\gamma^{(2)}(u,v;u',v') \,.
\eeqn
For a function $a:\R^n\times\R^n\rightarrow\R$, $(x,y)\mapsto a(x,y)$,
we define the Morawetz action
\eqn
	M_a \, := \, \int dx \, dy \, \nabla_x a(x,y) \cdot P_x(x,y)   \,.
\eeqn
where
\eqn
	P_x(x,y) & := &
	\int du \, du' \, dv \, dv' \,  e^{i(u-u')x+i(v-v')y}\left(\frac{u+u'}2\right) \,
	\widehat\gamma^{(2)}(u,v;u',v')
\eeqn
so that
\eqn
	P(x) \, = \, \int dy \, P_x(x,y) \,.
\eeqn
The time derivative is given by
\eqn\label{eq-dtMa-terms-1}
	\partial_t M_a & = & \int dx \, dy \,  \nabla_x a(x,y) \cdot \partial_t P_x(x,y)
	 \nonumber\\
	 &=&\frac12 \, \Big[ \,  (A_1) \, + \, (A_2) \, + \, (A_3) \, + \, (A_4) \, \Big] \,,
\eeqn
where the four terms on the rhs are defined as follows.

We have
\eqn
	(A_1) & := & \frac1i\int dx \, dy \, \big( \, \nabla_x a(x,y)
	\cdot\, \int du \, du' \, dv \, dv' \, e^{i(u-u')x+i(v-v')y} \,
	\nonumber\\
	&&\hspace{1cm} (u+u') \,(u^2 - (u')^2)  \, \widehat\gamma^{(2)}(u,v;u',v')
\eeqn
and
\eqn
	(A_2) &: = & \frac1i\int dx \, dy \, \big( \, \nabla_x a(x,y)
	\cdot\, \int du \, du' \, dv \, dv' \, e^{i(u-u')x+i(v-v')y} \,
	\nonumber\\
	&&\hspace{1cm}
	(u+u') \,(v^2 - (v')^2)
	\, \widehat\gamma^{(2)}(u,v;u',v') \,.
\eeqn
Moreover,
\eqn
	(A_3) & := & \frac\lambda i\int dx \, dy \, \big( \, \nabla_x a(x,y)
	\cdot\, \int du \, du' \, dv \, dv' \, e^{i(u-u')x+i(v-v')y} \,
	\nonumber\\
	&&\hspace{1cm} (u+u') \,
	 \big( \widehat{B^+_{1,3}\gamma^{(3)}}(u,v;u',v')
	\, - \,  \widehat{B^-_{1,3}\gamma^{(3)}}(u,v;u',v') \big)
\eeqn
and
\eqn
	(A_4) &: = & \frac\lambda i\int dx \, dy \, \big( \, \nabla_x a(x,y)
	\cdot\, \int du \, du' \, dv \, dv' \, e^{i(u-u')x+i(v-v')y} \,
	\nonumber\\
	&&\hspace{1cm}
	(u+u') \,
	\, \big( \widehat{B^+_{2,3}\gamma^{(3)}}(u,v;u',v')
	\, - \,  \widehat{B^-_{2,3}\gamma^{(3)}}(u,v;u',v') \big) \,.
\eeqn
We now discuss each of these four terms in detail.

\subsubsection{The term $(A_1)$}
We have
\eqn
	\lefteqn{
	(A_{1})
	}
	\nonumber\\
	& = & \frac1i\int dx \, dy\, \nabla_x a(x,y)
	\cdot\int du \, du' \,dv \, dv' \,
	e^{i(u-u')x+i(v-v')y} \,  (u+u') \, (u^2 - (u')^2) \,
	\nonumber\\
	&&\hspace{4cm}
	\widehat\gamma^{(2)}(u,v;u',v')
	\nonumber\\
	& = & \frac1i\int dx \, dy \, \nabla_x a(x,y)
	\cdot\int du \, du' \,  dv \, dv' \,
	\widehat\gamma^{(2)}(u,v;u',v') \,
	\nonumber\\
	&&\hspace{4cm}   [(u+u')\otimes(u+u')](u-u') \,  e^{i(u-u')x+i(v-v')y}
	\nonumber\\
	& = & -\int dx \, dy \, \int du \, du' \, dv \, dv' \,   \widehat\gamma^{(2)}(u,v;u',v') \,
	\nonumber\\
	&&\hspace{4cm}
	\nabla_x a(x,y) \cdot [(u+u')\otimes(u+u')] (\nabla_x  e^{i(u-u')x+i(v-v')y} \,   )\,
	\nonumber\\
	& = &  \int dx \, dy \int du \, du' \, dv \, dv' \, \widehat\gamma^{(2)}(u,v;u',v') \,\\
	&&\hspace{4cm}
	\sum_{i,j}(\partial_{x_i}\partial_{x_j} a(x,y)) \, \tu_i\tu_j \,    e^{i(u-u')x+i(v-v')y} \,
	\nonumber
\eeqn
where $\tu:=u+u'$.
This equals
\eqn
	& = &   \int dx dx' \, dy \, \delta(x-x') \, \int du \, du' \, dv\, dv' \,
	\sum_{i,j}(\partial_{x_i}\partial_{x_j} a(x,y)) \,
	\nonumber\\
	&&\hspace{5.5cm}  \tu_i\tu_j \,   e^{i(ux-u'x')+i(v-v')y} \,
	\widehat\gamma^{(2)}(u,v;u',v')
	\nonumber\\
	& = & -\, \int dx dx' \, dy \, \delta(x-x') \,
	\sum_{i,j}(\partial_{x_i}\partial_{x_j} a(x,y)) \,
	(\partial_{x_i}-\partial_{x_i'})(\partial_{x_j}-\partial_{x_j'}) \, \gamma^{(2)}(x,y;x',y)
	\nonumber\\
	& = & -\, \int dx dx' \, dy \, \delta(x-x') \,
	\sum_{i,j}(\partial_{x_i}\partial_{x_j} a(x,y)) \,
	(\partial_{x_i}\partial_{x_j}+\partial_{x_i'} \partial_{x_j'}) \, \gamma^{(2)}(x,y;x',y)
	\nonumber\\
	& & + \,  \int dx dx' \, dy \, \delta(x-x') \,
	\sum_{i,j}(\partial_{x_i}\partial_{x_j} a(x,y)) \,
	(\partial_{x_i'}\partial_{x_j}+\partial_{x_i} \partial_{x_j'} ) \, \gamma^{(2)}(x,y;x',y)
	\nonumber\\
	& = & -\, \int dx dx' \, dy \, \delta(x-x') \,
	\sum_{i,j}(\partial_{x_i}\partial_{x_j} a(x,y)) \,
	\nonumber\\
	&&\hspace{5.5cm}
	(\partial_{x_i}\partial_{x_j}+
	\partial_{x_i}\partial_{x_j'}+\partial_{x_i'} \partial_{x_j}+\partial_{x_i'} \partial_{x_j'})
	\, \gamma^{(2)}(x,y;x',y)
	\nonumber\\
	& & + \, 2 \int dx dx' \, dy \, \delta(x-x') \,
	\sum_{i,j}(\partial_{x_i}\partial_{x_j} a(x,y)) \,
	(\partial_{x_i'}\partial_{x_j}+\partial_{x_i} \partial_{x_j'}) \, \gamma^{(2)}(x,y;x',y)
	\nonumber\\
	& = & -\, \int dx dy \,
	\sum_{i,j}(\partial_{x_i}\partial_{x_j} a(x,y)) \,  \partial_{x_i}\partial_{x_j} \, \gamma^{(2)}(x,y;x,y)
	\nonumber\\
	& & + \, 4\Re\int dx dx' \, dy \, \delta(x-x') \,
	\sum_{i,j}(\partial_{x_i}\partial_{x_j} a(x,y)) \,  \partial_{x_i} \partial_{x_j'}  \, \gamma^{(2)}(x,y;x',y)
\eeqn
where we used $\gamma^{(2)}(x,y;x',y)=\overline{\gamma^{(2)}(x',y;x,y)}$, and applied a coordinate
change $x\leftrightarrow x'$ in one of the two integrals contributing to the last line.
This equals
\eqn
	& = &- \, \int dx dx' \, dy \,
	(\Delta_xa(x,y))   \, \Delta_x \gamma^{(2)}(x,y;x,y)
	\nonumber\\
	& & + \, 4\Re\int dx dx' \, dy \, \delta(x-x') \,
	\sum_{i,j}(\partial_{x_i}\partial_{x_j} a(x,y)) \,  \partial_{x_i} \partial_{x_j'}  \, \gamma^{(2)}(x,y;x',y) \,.
\eeqn
This corresponds to the first and last term on the rhs of (3.36) in \cite{tt}.

\subsubsection{The term $(A_2)$}

We have
\eqn
	(A_2) &: = & \frac1i\int dx \, dy \, \big( \, \nabla_x a(x,y) \, \big)
	\cdot\, \int du \, du' \, dv \, dv' \, e^{i(u-u')x+i(v-v')y} \,
	\nonumber\\
	&&\hspace{1cm}
	[(u+u') \otimes (v+v')](v - v')
	\, \widehat\gamma^{(2)}(u,v;u',v') \nonumber\\
	&=&- \int dx \, dy \, \sum_{j,\ell}\big( \, \partial_{x_j} a(x,y) \, \big) \cdot
	\, \int du \, du' \, dv \, dv' \, (\partial_{y_\ell}  e^{i(u-u')x+i(v-v')y} )\,
	\nonumber\\
	&&\hspace{1cm}
	(u+u')_j   (v+v')_{\ell}
	\, \widehat\gamma^{(2)}(u,v;u',v')
	\nonumber\\
	&=& \int dx \, dy \, \big( \, \partial_{x_j} \partial_{y_\ell}  a(x,y) \, \big)
	\\
	&&\hspace{1cm}
	\, \int du \, du' \, dv \, dv' \, e^{i(u-u')x+i(v-v')y} \,
	(u+u')_j   (v+v')_{\ell}
	\, \widehat\gamma^{(2)}(u,v;u',v') \,.
	\nonumber
\eeqn
Here we note that if $\gamma^{(2)}(x,y;x',y')=\gamma^{(1)}(x;x')\gamma^{(1)}(y;y')$ has product form,
then the integral on the last line corresponds to $4P(x)P(y)$, the product of momentum densities
defined in \eqref{eq-Pmom-def-1}.

\subsubsection{The term $(A_3)$}

We have
\eqn
	\lefteqn{
	B_{1,3}^+\gamma^{(3)}(x,y;x',y')
	}
	\nonumber\\
	& = &
	\int dz \, dz' \, \delta(x-z) \, \delta(x-z') \,
	\nonumber\\
	&&\quad\quad
	\int du dv dq du' dv' dq'
	e^{i(ux+vy+qz-u'x'-v'y'-q'z')} \, \widehat\gamma^{(3)}(u,v,q;u',v',q')
	\nonumber\\
	 & = &
	\int du dv dq du' dv' dq'
	e^{i((u+q-q')x-u'x'+vy-v'y')} \, \widehat\gamma^{(3)}(u,v,q;u',v',q')
	\,.
\eeqn
Therefore,
\eqn
	\lefteqn{
	\widehat{B_{1,3}^+\gamma^{(3)}}(u,v;u',v')
	}
	\nonumber\\
	& = & \int dxdx' dy dy' \, e^{-iux-ivy+iu'x'+iv'y'} \, (B_{1,3}^+\gamma^{(3)})(x,y;x',y')
	\nonumber\\
	& = &
	\int dqdq'  \,  \widehat\gamma^{(3)}(u-q+q',v,q;u',v',q') \,.
\eeqn
Likewise, one obtains
\eqn
	\widehat{B_{1,3}^-\gamma^{(3)}}(u,v;u',v')
	& = &
	\int dqdq'  \,  \widehat\gamma^{(3)}(u,v,q;u'+q-q',v',q') \,.
\eeqn

Now, in order to consider $(A_3)$ we first look at
\eqn
	\lefteqn{
	\frac1i\int dudu'dvdv'  \, e^{i(u-u')x+i(v-v')y} \, (u+u') \,
	(\widehat{B_{1,3}^+\gamma^{(3)}}(u,v;u',v') -
	\widehat{B_{1,3}^-\gamma^{(3)}}(u,v;u',v'))
	}
	\nonumber\\
	& = &
	\frac1i\int dudu'dvdv' dqdq' \, e^{i(u-u')x+i(v-v')y} \, (u+u') \, \widehat\gamma^{(3)}(u-q+q',v,q;u',v',q')
	\nonumber\\
	&&- \, \frac1i\int dudu'dvdv' dqdq' \, e^{i(u-u')x+i(v-v')y} \,
	(u+u') \, \widehat\gamma^{(3)}(u,v,q;u'+q-q',v',q')   \,.
	\quad\quad
\eeqn
In the last term, we apply the change of variables $u\rightarrow u-q+q'$ and $u'\rightarrow u'-q+q'$,
so that the difference $u-u'$ remains unchanged.
We obtain that the above equals
\eqn
	\lefteqn{
	\frac1i\int dudu'dvdv' dqdq' \, e^{i(u-u')x+i(v-v')y} \,  (u+u') \, \widehat\gamma^{(3)}(u-q+q',v,q;u',v',q')
	}
	\nonumber\\
	&&- \, \frac1i\int dvdv' dqdq' \, e^{i(u-u')x+i(v-v')y} \,
	(u+u'-2q+2q') \, \widehat\gamma^{(3)}(u-q+q',v,q;u',v',q')
	\nonumber\\
	&=&\frac1i\int dudu'dvdv' dqdq' \, e^{i(u-u')x+i(v-v')y}
	\, \big( \, (u+u') \, - \,  (u+u'-2q+2q') \, \big)
	\nonumber\\
	&&\hspace{7cm}
	\,  \widehat\gamma^{(3)}(u-q+q',v,q;u',v',q')
	\nonumber\\
	&=&\frac1i\int dvdv' dqdq' \, e^{i(u-u')x+i(v-v')y} \,
	2(q-q') \, \widehat\gamma^{(3)}(u-q+q',v,q;u',v',q') \,.
\eeqn
The contribution of this term to $(A_3)$ is given by
\eqn
	\lefteqn{
	\lambda \frac1i\int dx dy \nabla_x a(x,y)\cdot \int dudu'dvdv' dqdq' \, e^{i(u-u')x+i(v-v')y} \,
	}
	\nonumber\\
	&&\hspace{4cm}
	2(q-q') \,
	\widehat\gamma^{(3)}(u-q+q',v,q;u',v',q') \,.
\eeqn
Next, we express everything in position space.

We have that the last line equals
\eqn
	\lefteqn{
	\frac{\lambda}{i}\int dx dy \; \nabla_x a(x,y)
	\cdot \int dXdYdZdX'dY'dZ' \int dudu'dvdv' dqdq' \, e^{i(u-u')x+(v-v')y} \, 2(q-q') \,
	}
	\nonumber\\
	&&\quad\quad\quad\quad
	e^{i(-(u-q+q')X-vY-qZ+u'X'+v'Y'+q'Z')} \gamma^{(3)}(X,Y,Z;X',Y',Z')
	\nonumber\\
	&=&
	\frac{\lambda}{i}\int dx dy   \int dXdYdZdX'dY'dZ' \,
	\gamma^{(3)}(X,Y,Z;X',Y',Z')
	\, \int dudu'dvdv' dqdq' \,
	\nonumber\\
	&&\quad\quad\quad\quad
	e^{iu(x-X)+iv(y-Y)-iu'(x-X')-iv'(y-Y')} \, 2\nabla_x a(x,y)\cdot(q-q') \, e^{+iq(X-Z)-q'(X-Z')}\,   \,
	\nonumber\\
	&=&-
	\lambda \int dx dy  \int dXdYdZdX'dY'dZ'\,
	\gamma^{(3)}(X,Y,Z;X',Y',Z') \int dqdq' \,
	\nonumber\\
	&&\quad\quad\quad\quad
	\delta(x-X)\delta(x-X') \,
	\delta(y-Y)\delta(y-Y') \,
	2\nabla_x a(x,y)\cdot\nabla_X \, e^{+iq(X-Z)-iq'(X-Z')}\,
	\nonumber\\
	&=&
	-\lambda \int dXdYdZdZ' \,
	\gamma^{(3)}(X,Y,Z;X ,Y,Z')
	\nonumber\\
	&&\quad\quad\quad\quad
	\, 2\nabla_X a(X,Y)\cdot\nabla_X \, \delta(X-Z) \, \delta(Z-Z') \,
	\label{eq-nonlinterm-aux-1-1}\\
	&=&
	-\lambda \int dXdYdZ  \,
	\gamma^{(3)}(X,Y,Z;X ,Y,Z)
	\, 2\nabla_X a(X,Y)\cdot\nabla_X \, \delta(X-Z)
	\nonumber\\
	&=&
	\lambda \int dXdYdZ  \, \delta(X-Z) \, (2\Delta_X a(X,Y)+ 2(\nabla_X a(X,Y))\cdot\nabla_X \,  )
	\nonumber\\
	&&\hspace{8cm}
	\gamma^{(3)}(X,Y,Z;X ,Y,Z)
	\label{eq-nonlinterm-aux-2-1}
\eeqn
where we have written $\delta(X-Z)\delta(X-Z')=\delta(X-Z)\delta(Z-Z')$
to get (\ref{eq-nonlinterm-aux-1-1}).

Now we rename the variables $(X,Y,Z)\rightarrow(x,y,z)$, and note that
\eqn
	\lefteqn{
	\int dx dy \,
	(\nabla_x a(x,y))\cdot\nabla_x\gamma^{(3)}(x,y,x;x,y,x)
	}
	\nonumber\\
	&=&
	\int dxdydz  \, \delta(x-z) \, ( \,  (\nabla_x a(x,y))\cdot\nabla_x
	\,  + \, (\nabla_z a(z,y))\cdot\nabla_z \, )
	\gamma^{(3)}(x,y,z;x ,y,z)
	\nonumber\\
	&=&
	\int dxdydz  \, \delta(x-z) \, ( \, (\nabla_x a(x,y))\cdot\nabla_x \gamma^{(3)}(x,y,z;x ,y,z)
	\nonumber\\
	&&\hspace{7cm}
	\,  + \, (\nabla_z a(z,y))\cdot\nabla_z \gamma^{(3)}(z,y,x;z,y,x) \, )
	\nonumber\\
	&=&
	\int dxdydz  \, \delta(x-z) \, ( \,  2(\nabla_x a(x,y))\cdot\nabla_x  \gamma^{(3)}(x,y,z;x ,y,z)  \, )
	\label{eq-cubic-div-1}
\eeqn
where we used the symmetry $\gamma^{(3)}(x,y,z;x,y,z)=\gamma^{(3)}(z,y,x;z,y,x)$,
and renamed the variables in the last term.
Clearly, the left hand side equals
\eqn\label{eq-cubic-dim-1}
	- \int dx dy \, (\Delta_x a(x,y)) \, \gamma^{(3)}(x,y,x;x,y,x)
\eeqn
from integrating by parts.

Therefore, combining \eqref{eq-nonlinterm-aux-2-1},  \eqref{eq-cubic-div-1} and
\eqref{eq-cubic-dim-1}
\eqn
	(A_3)
	& = & \lambda \int dxdydz  \, \delta(x-z) \, (2\Delta_x a(x,y)-\Delta_x a(x,y) \,  )
	\gamma^{(3)}(x,y,z;x ,y,z)
	\nonumber\\
	& = & \lambda \int dxdy \, (\Delta_x a(x,y))
	\gamma^{(3)}(x,y,x;x,y ,x) \,.
\eeqn
This corresponds to the second term on the rhs of (3.36) in \cite{tt}.

\subsubsection{The term $(A_4)$}

We have
\eqn
	(A_4) &= & \frac\lambda i\int dx \, dy \,  \nabla_x a(x,y)
	\cdot\, \int du \, du' \, dv \, dv' \, e^{i(u-u')x+i(v-v')y} \,
	\nonumber\\
	&&\hspace{1cm}
	(u+u') \,
	\, \big[ \widehat{B^+_{2,3}\gamma^{(3)}}(u,v;u',v')
	\, - \,  \widehat{B^-_{2,3}\gamma^{(3)}}(u,v;u',v') \big]
	\nonumber\\
	&=:&\frac\lambda i\int dx \, dy \,  \nabla_x a(x,y)
	\cdot\, \int du \, du' \, dv \, dv' \, dq \, dq' \, e^{i(u-u')x+i(v-v')y} \,
	\label{eq-A4-auxid-1}\\
	&&\hspace{1cm}
	(u+u') \,
	\, \big[ \widehat{\gamma^{(3)}}(u,v-q+q',q;u',v',q')
	\, - \, \widehat{\gamma^{(3)}}(u,v,q;u',v'+q-q',q') \big]
	\nonumber\\
	&=:&\frac\lambda i\int dx \, dy \,   \nabla_x a(x,y)
	\cdot\, \int du \, du' \, dv \, dv' \, dq \, dq' \, e^{i(u-u')x+i(v-v')y} \,
	\label{eq-A4-auxid-2}\\
	&&\hspace{1cm}
	(u+u') \,
	\, \big[ \widehat{\gamma^{(3)}}(u,v+q',q;u',v'+q,q')
	\, - \, \widehat{\gamma^{(3)}}(u,v+q',q;u',v'+q,q') \big]
	\nonumber\\
	&=&0
\eeqn
where to pass to \eqref{eq-A4-auxid-2},
we used the coordinate change $v\rightarrow v+q$, $v'\rightarrow v'+q$ for the
expression involving the first term in the square bracket in \eqref{eq-A4-auxid-1},
and  $v\rightarrow v+q'$, $v'\rightarrow v'+q'$ for the second term.
Both coordinate changes leave the difference $v-v'$ invariant.

\subsubsection{Completing the proof}
Summarizing, we obtain from \eqref{eq-dtMa-terms-1} that the following result holds.

\begin{theorem}
\label{thm-Morawetz-main-1}
Under the conditions formulated above, the {\em interaction Morawetz identity}
\eqn
	\partial_t M_a
	&=& - \, \frac12\int dx dy \,
	(\Delta_x a(x,y))   \, \Delta_x\gamma^{(2)}(x,y;x,y)
	\nonumber\\
	&&+ \, \frac \lambda2 \int dxdy \, (\Delta_x a(x,y))
	\gamma^{(3)}(x,y,x;x,y,x)
	\nonumber\\
	&&+ \, 2\Re \int dx dx' dy \, \delta(x-x') \,
	\sum_{j,\ell} \, \big( \, \partial_{x_j}  \partial_{x_\ell} a(x,y) \, \big)
	\partial_{x_\ell}\partial_{x_j'}
	\gamma^{(2)}(x,y;x',y)
	\nonumber\\
	&&+ \, 2\int dx \, dy \, \sum_{j,\ell} \, \big( \, \partial_{x_j} \, \partial_{y_\ell}  a(x,y) \, \big)
	\, \int du \, du' \, dv \, dv' \, e^{i(u-u')x+i(v-v')y} \,
	\nonumber\\
	&&\hspace{4cm}
	\left(\frac{u+u'}2\right)_j \;  \left(\frac{v+v'}2\right)_{\ell}
	\, \widehat\gamma^{(2)}(u,v;u',v')
	\,.
	\label{eq-intMorawetz-final-1}
\eeqn
holds for solutions of cubic GP hierarchies.
\end{theorem}

While we are not invoking any particular choice of $a(x,y)$ in this paper, we
remark that there exists $a$ such that $\Delta \Delta a = \delta(x_1 - x_2) \delta(x_1 - x_3)$. 
For details on this issue, we refer to \cite{xch-die} and \cite{gm-jhde}.  

Now we compare the result of Theorem \ref{thm-Morawetz-main-1}
with \eqref{mainestimate} obtained above for the NLS.

For factorized solutions of the cubic GP hierarchy of the form
\eqref{eq-gammak-factorized-1}, we obtain the following:
\eqn\label{eq-GPinMorawetz-final-1}
	\partial_t M_a
	&=&- \, \frac12 \, \int dx dy \,
	|\phi(y)|^2 \,
	(\Delta_x a(x,y))   \, \Delta_x\rho(t,x)
	\nonumber\\
	&&+ \,  \frac\lambda2 \int dx \, (\Delta_x a(x,y)) \, |\phi(t,y)|^2 \, |\phi(t,x)|^4
	\nonumber\\
	&&+ \, 2 \, \Re \int dx dy \,  |\phi(y)|^2 \,
	\sum_{j,\ell} \, \big( \, \partial_{x_j}  \partial_{x_\ell} a(x,y) \, \big)
	(\partial_{x_\ell}\phi(t,x)) \, \overline{(\partial_{x_j}  \phi(t,x))} \,
	\nonumber\\
	&&+ \, 2 \int dx \, dy \, \sum_{j,\ell} \,
	\big( \, \partial_{x_j} \, \partial_{y_\ell}  a(x,y) \, \big) (P_\phi(t,x))_j \;
	(\overline{P_\phi(t,y)})_\ell  \,     \,,
\eeqn
where
\eqn	
	P_\phi(x) \, := \, \Im( \, \overline{\phi(x)}\nabla_x\phi(x) \,)
\eeqn
is the momentum density corresponding to $\phi(t,x)$.
This corresponds to \eqref{mainestimate} for the cubic NLS where $p=3$ (so that
$\lambda\frac{p-1}{p+1}=\frac\lambda2$).
In particular, we note that for $a(x,y)=\widetilde a(x-y)$, we evidently have
$\partial_{x_j} \partial_{y_\ell}  a(x,y)=-\partial_{x_j}  \partial_{x_\ell}  a(x,y)$,
which agrees with \eqref{mainestimate}.

\subsection{Proof of the one-particle Morawetz identities}
\label{ssec-simpMorawetz-1}

In this section, we prove the standard (single-particle) Morawetz identities
in Theorem \ref{thm-Morawetz-main-2},
as a corollary of the interaction Morawetz identities derived above.

The corresponding explicit expression for
$\partial_t M_a$ can be easily obtained from the interaction
Morawetz identities \eqref{eq-intMorawetz-final-1}, by choosing
\eqn
	a(x,y) \, = \, a(x)
\eeqn
independent of $y$, and
\eqn
	\gamma^{(1)}(x;x') & = & \int dy \, \gamma^{(2)}(x,y;x',y)
	\nonumber\\
	\gamma^{(2)}(x,z;x',z') & = & \int dy \, \gamma^{(3)}(x,y,z;x',y,z') \,,
\eeqn
which follows from the admissibility of the density matrices, see \eqref{eq-admiss-def-1}.

Accordingly, \eqref{eq-intMorawetz-final-1} reduces to
\eqn
	\partial_t M_a
	&=&-\frac{1}{2} \, \int dx \,
	(\Delta_x \Delta_x a(x))   \, \gamma^{(1)}(x;x)
	\nonumber\\
	&&+ \, \frac\lambda2 \int dx \, (\Delta_x a(x))
	\gamma^{(2)}(x,x;x,x)
	\nonumber\\
	&&+ 2 \, \Re \int dx dx' \delta(x-x') \,
	\sum_{j,\ell} \, \big( \, \partial_{x_j}  \partial_{x_\ell} a(x) \, \big)
	\partial_{x_\ell}\partial_{x_j'}
	\gamma^{(1)}(x;x')   \,.
	\label{eq-simpMorawetz-final-1}
\eeqn
We note that the term involving the momentum densities on the last
line of \eqref{eq-GPinMorawetz-final-1} is not present here (since $\partial_{y_\ell}a(x)=0$).

\subsection*{Acknowledgements} 
T.C. was supported by the NSF through grants DMS-1009448 and DMS-1151414 (CAREER). The work of N.P. was supported by NSF grants DMS-0758247 and DMS-1101192 and an Alfred P. Sloan Research Fellowship. The work of N.T. was supported by NSF grant DMS-0901222.

\end{document}